\documentclass[journal]{IEEEtran}

\usepackage{epsf,amsmath,amscd,amssymb,graphics,latexsym,multicol}
\def\IR{{\rm I \kern-0.20em R}}

\newtheorem{thm}{Theorem}

\newtheorem{defn}{Definition}
\newtheorem{lem}{Lemma}

\begin{document}
\title{Consistency in Models for Distributed Learning under Communication Constraints\footnote{This paper was presented in part at the 17th Annual Conference on Learning Theory (COLT), Banff, Canada, July 1-4, 2004 \cite{PreKulPoo04b} and in part at the 42nd Annual Allerton Conference on Communication, Control, and Computing, Monticello, IL, Sept 29-Oct 1, 2004 \cite{PreKulPoo04d}.}}
\author{Joel~B.~Predd,~\IEEEmembership{Member,~IEEE,}
        Sanjeev~R.~Kulkarni,~\IEEEmembership{Fellow,~IEEE,}
        and~H.~Vincent~Poor,~\IEEEmembership{Fellow,~IEEE}% <-this % stops a space
\thanks{This research was supported in part by the Army Research Office under grant
DAAD19-00-1-0466, in part by Draper Laboratory under grant IR\&D
6002, in part by the National Science Foundation under grant
CCR-0312413, and in part by the Office of Naval Research under Grant No.
N00014-03-1-0102.  }
\thanks{The authors are with the Department of Electrical Engineering, Princeton University, Princeton, NJ 08544 USA (email: jpredd/kulkarni/poor@princeton.edu)}}

\maketitle

\begin{abstract}
Motivated by sensor networks and other distributed settings, several models for distributed learning are presented.  The models differ from classical works in statistical pattern recognition by allocating observations of an independent and identically distributed (i.i.d.) sampling process amongst members of a network of simple learning agents.  The agents are limited in their ability to communicate to a central fusion center and thus, the amount of information available for use in classification or regression is constrained.  For several basic communication models in both the binary classification and regression frameworks, we question the existence of agent decision rules and fusion rules that result in a universally consistent ensemble; the answers to this question present new issues to consider with regard to universal consistency. This paper addresses the issue of whether or not the guarantees provided by Stone's Theorem in centralized environments hold in distributed settings.
\end{abstract}


\IEEEpeerreviewmaketitle

\section{Introduction}
\subsection{Models for Distributed Learning}
Consider the following learning model:  Let $X$ and $Y$ be $\mathcal{X}$-valued and $\mathcal{Y}$-valued random variables, respectively,  with a joint distribution denoted by $\mathbf{P}_{XY}$.   ${\mathcal{X}}$ is known as the feature, input, or observation space; ${\mathcal{Y}}$ is known as the label, output, or target space.   Throughout, we take ${\mathcal{X}}\subseteq\IR^d$ and consider two cases corresponding to binary classification (${\mathcal{Y}}=\{0,1\}$) and regression estimation (${\mathcal{Y}}=\IR$).  Given a loss function $l:{\mathcal{Y}}\times{\mathcal{Y}}\rightarrow\IR$, the decision-theoretic problem is to design a decision rule $g:{\cal{X}}\rightarrow{\cal{Y}}$ that achieves the minimal expected loss $L^{\star}=\inf_{g}{\mathbf{E}}\{l(g(X), Y)\}$.  Without prior knowledge of the distribution ${\mathbf{P}}_{XY}$, computing a loss minimizing decision rule is not possible.  Instead,  $D_n=\{(X_i, Y_i)\}_{i=1}^{n}$, an independent and identically distributed (i.i.d.) collection of training data with $(X_i, Y_i)\sim \mathbf{P}_{XY}$ for all $i\in\{1,...,n\}$ is available;  the learning problem is to use this data to infer decision rules with small expected loss.  

This standard learning model invites one to consider numerous questions; however in this work, we focus on the statistical property known as \emph{universal consistency} \cite{DevGyoLug96,GyoKohKrzWal02}. In traditional, centralized settings,  $D_n$ is provided to a single learning agent, and questions have been posed about the existence of classifiers or estimators that are universally consistent. The answers to such questions are well understood and are provided by results such as Stone's Theorem \cite{Sto77}, \cite{DevGyoLug96}, \cite{GyoKohKrzWal02} and numerous others in the literature.  

Suppose, in contrast with the standard centralized setting, that for each $i\in\{1,...,n\}$, the training datum $(X_i, Y_i)$ is received by a distinct member of a network of $n$ simple learning agents.    When a central authority observes a new observation $X\sim\mathbf{P}_X$,  it broadcasts the observation to the network in a request for information.   At this time, each agent can respond with at most one bit.  That is, each learning agent chooses whether or not to respond to the central authority's request for information; if it chooses to respond, an agent sends either a $1$ or a $0$ based on its local decision algorithm.  Upon observing the response of the network, the central authority acts as a fusion center, combining the information to create an estimate of $Y$.  As in the centralized setting, a key question arises:  do there exist agent decision rules and a fusion rule that result in a universally consistent network in the limit as the number of agents increases without bound?

In what follows, we answer this question in the affirmative for both binary classification and regression estimation.   In the binary classification setting, we demonstrate agent decision rules and a fusion rule that correspond nicely with classical kernel classifiers.  With this connection to classical work, the universal Bayes-risk consistency of this ensemble then follows immediately from celebrated analyses like Stone's Theorem, etc.  In the regression setting, we demonstrate that under regularity, randomized agent decision rules exist such that when the central authority applies a scaled average vote combination of the agents' responses,  the resulting estimator is universally consistent under $L_2$-loss.   

In this model, the agents convey slightly more information than is suggested by the mere one bit that we have allowed them to physically transmit to the fusion center.  Indeed, each agent decides not between sending $1$ or $0$.  Rather, each agent's decision rule can be viewed as a selection of one of \textit{three} states:  abstain, vote and send $0$, and vote and send $1$.  With this observation, these results can be interpreted as follows: $\log_2(3)$ bits per agent per classification is sufficient for universal consistency to hold for both distributed classification and regression \textit{with abstention}.

In this view, it is natural to ask whether these $\log_2(3)$ bits are necessary.  Can consistency results be proven at lower bit rates?  Consider a revised model, precisely the same as above, except that in response to the central authority's request for information, each agent must respond with 1 or 0;  abstention is not an option and thus, each agent responds with exactly one bit per classification. Are there rules for which universal consistency results hold in distributed classification and regression \textit{without abstention}?

Interestingly, we demonstrate that in the binary classification setting,  randomized agent decision rules exist such that when a majority vote fusion rule is applied, universal Bayes-risk consistency holds.  Next, we establish natural regularity conditions for candidate fusion rules and specify a reasonable class of agent decision rules. As an important negative result, we then demonstrate that for any agent decision rule within the class, there does not exist a regular fusion rule that is $L_2$ consistent for every distribution ${\mathbf{P}}_{XY}$.  This result establishes the impossibility of universal consistency in this model for distributed regression without abstention for a restricted, but reasonable class of decision rules.
 
 \subsection{Motivation and Background}
Motivation for studying distributed learning in general and the current models in particular arise from wireless sensor networks and distributed databases, applications that have attracted considerable attention in
recent years \cite{AkySuSanCay02}.  Research in wireless sensor networks has focused on two separate
aspects:  networking issues, such as capacity,
delay, and routing strategies; and applications issues.  This paper
is concerned with the second of these aspects, and in particular with the problem of distributed inference.
Wireless sensor networks are {\it a fortiori} designed for the purpose
of making inferences about  the environments that they are sensing,
and they are typically characterized by limited communications
capabilities due to tight energy and bandwidth limitations, as well as the typically ad-hoc nature of wireless networks. Thus, distributed inference is a major issue in the study of wireless sensor networks.

In problems of distributed databases, there is a collection of training data that is massive in both the dimension of the feature space and quantity of data. For political, economic, social or technological reasons, this database is distributed geographically or in such a way that it is infeasible for any single agent to access the entire database.  Multiple agents may be deployed to make inferences from various segments of the database, but communication constraints arising from privacy or security concerns highlight distributed inference as a key issue in this setting as well.  Recent research has studied inference in the distributed databases setting from an algorithmic point of view; for example, \cite{LazObr01} proposed a distributed boosting algorithm and studied its performance empirically.

Distributed detection and estimation is a well-developed field
with a rich history.  Much of the work in this area has focused on
either parametric problems, in which strong statistical assumptions
are made \cite{Tsi93, Var96, BluKasPoo97,Vee01, LiWonHuSay02, KumZhaShe02, CosAay03, KotRamSay03, DonTonSad04}, or on traditional nonparametric formalisms, such as
constant-false-alarm-rate detection \cite{BarVar89}. Recently, \cite{Sim03} advocated a learning theoretic approach to wireless sensor networks and \cite{NguWaiJor04}, in the context of kernel methods commonly used in machine learning, considered the classical model for decentralized detection \cite{Tsi93} in a nonparametric setting.     

In this paper, we consider an alternative nonparametric approach to the study of distributed inference that is most closely aligned with models considered in nonparametric statistics and the study of kernel estimators and other Stone-type rules. Extensive work has been done related to the consistency of Stone-type rules under various sampling processes; for example, \cite{DevGyoLug96}, \cite{GyoKohKrzWal02} and references therein, \cite{Cov68},  \cite{GrePaw87}, \cite{Krz86}, \cite{KulPos95}, \cite{KulPosSan02}, \cite{MorKulNob99}, \cite{Nob99}, \cite{NobAda01}, \cite{NobMorKul98}, \cite{Rou91}, \cite{Sto77}, \cite{Yak89}, \cite{Yak93}.  These models focus on various dependency structures within the training data and assume that a single processor has access to the entire data stream.  

The nature of the work considered in this paper is to consider similar questions of universal consistency in models that capture some of the structure in a distributed environment.    As motivated earlier, agents in distributed scenarios have constrained communication capabilities and moreover, each may have access to distinct data streams that differ in distribution and may depend on parameters such as the state of a sensor network or location of a database.  We consider the question: for a given model of communication amongst agents, each of whom has been allocated a small portion of a larger learning problem,  can enough information can be exchanged to allow for a universally consistent ensemble?  In this work, the learning problem is divided amongst agents by allocating each a unique observation of an i.i.d. sampling process.  As explained earlier, we consider simple communication models with and without abstention.  Insofar as these models present a useful picture of distributed scenarios, this paper addresses the issue of whether or not the guarantees provided by Stone's Theorem in centralized environments hold in distributed settings.  Notably, the models under consideration will be similar in spirit to their classical counterparts;  indeed, similar techniques can be applied to prove results.  

Note that \cite{PreKulPoo04a} studies a similar model for distributed learning under communication constraints.  Whereas \cite{PreKulPoo04a} allocates regions of feature space amongst agents, here we allocate observations of an i.i.d. sampling process.  Moreover, here we study a richer class of communication constraints.  A related area of research lies in the study of ensemble methods in machine learning; examples of these techniques include bagging, boosting, mixtures of experts, and others \cite{JacJorNowHin91, Bre96, FreSch97b, FreSchSinWar97a, KitHatDuiMat98}.  These techniques are similar to the problem of interest here in that they aggregate many individually trained classifiers. However, the focus of these works is on the statistical and algorithmic advantages of learning with an ensemble and not on the nature of learning under communication constraints.  Notably, \cite{KeaSeu95} considered an PAC-like model for learning with many individually trained hypotheses in a distribution-specific (i.e., parametric) framework.

Numerous other works in the literature are relevant to the research presented here.  However, different points need to be made depending on whether we consider regression or classification with or without abstention. Lacking such context here, we will save such discussion of these results for the appropriate sections in the paper.  

\subsection{Organization}
The remainder of this paper is organized as follows.  In Section II, the notation and technical assumptions relevant to the remainder of the paper are introduced.  In Sections III and IV, we study the models for binary classification in communication with and without abstention, respectively.  In Sections V and VI, we study the models for regression estimation with and without abstention in turn.  In each section, we present the main results, discuss important connections to other work in nonparametric statistics, and then proceed with a proof that further emphasizes differences from classical analyses like Stone's Theorem.  In Section VII, we conclude with a discussion of future work.  Technical lemmas that are readily apparent from the literature are left to the appendix.

\section{Preliminaries}
In this section, we introduce notation and technical assumptions relevant to the remainder of the paper.

As stated earlier, let $X$ and $Y$ be $\mathcal{X}$-valued and $\mathcal{Y}$-valued random variables, respectively,  with a joint distribution denoted by $\mathbf{P}_{XY}$.   ${\mathcal{X}}$ is known as the feature, input, or observation space; ${\mathcal{Y}}$ is known as the label, output, or target space.   Throughout, we will take ${\mathcal{X}}\subseteq\IR^d$ and consider two cases corresponding to binary classification (${\mathcal{Y}}=\{0,1\}$) and regression estimation (${\mathcal{Y}}=\IR$).  Let $D_n=\{(X_i, Y_i)\}_{i=1}^{n}$ denote an i.i.d. collection of training data with $(X_i, Y_i)\sim \mathbf{P}_{XY}$ for all $i\in\{1,...,n\}$. 

Throughout this paper, we will use $\delta_{ni}$ to denote the randomized response of the $i^{th}$ learning agent in an ensemble of $n$ agents.  For each $i\in\{1,..., n\}$, $\delta_{ni}$ is an ${\mathcal{S}}$-valued random variable,    where $\mathcal{S}$ is the decision space for the agent;  in models \textit{with abstention} we take $\mathcal{S}=\{{\rm abstain}, 1,  0\}$ and in models \textit{without abstention} we take $\mathcal{S}=\{ 1,  0\}$.  As an important consequence of the assumed lack of inter-agent communication and the assumption that $D_n$ is i.i.d., we have the following observation which will be fundamental to the subsequent analysis:

\begin{description}
\item[(A)] The $i^{th}$ agent's response, $\delta_{ni}$, may be dependent on $X, X_i$, and $Y_i$, but is statistically independent of $\{(X_j, Y_j)\}_{j\neq i}$ and conditionally independent of $\{\delta_{nj}\}_{j\neq i}$ given $X$.
\end{description}

Thus, to specify $\delta_{ni}$ and thereby design agent decision rules, it suffices to define the conditional distribution ${\mathbf{P}}\{\delta_{ni}\,|X, X_i, Y_i\}$  for all $(X, X_i, Y_i)\in{\mathcal{X}}\times{\mathcal{X}}\times{\mathcal{Y}}$.  In each of the subsequent sections, we will find it convenient to do so by specifying a function $\bar{\delta}_{n}(x):{\mathcal{X}}\times{\mathcal{X}}\times\mathcal{Y}\rightarrow\{{\rm abstain}\}\cup [0,1]$.  In particular, we define
\begin{eqnarray}\nonumber
\lefteqn{{\mathbf{P}}\{\delta_{ni} = {\rm abstain}\,|X, X_i, Y_i\}} \\
\nonumber & =  &\left\{%
\begin{array}{ll}
    1, & {\rm if\,\,} \bar{\delta}_{n}(X, X_i, Y_i) = {\rm abstain} \\
    0, & {\rm otherwise}
\end{array}%
\right.\\
\label{randomizedrule}
\lefteqn{{\mathbf{P}}\{\delta_{ni} = 1\,|X, X_i, Y_i\}}\\
\nonumber &= &\left\{%
\begin{array}{ll}
    0, & {\rm if\,\,} \bar{\delta}_{n}(X, X_i, Y_i) = {\rm abstain} \\
    \bar{\delta}_{n}(X, X_i, Y_i), & {\rm otherwise}
\end{array}%
\right.\\
\nonumber
\lefteqn{{\mathbf{P}}\{\delta_{ni} = 0\,|X, X_i, Y_i\}}\\
\nonumber &=& \left\{%
\begin{array}{ll}
    0, & {\rm if\,\,} \bar{\delta}_{n}(X, X_i, Y_i) = {\rm abstain} \\
    1 - \bar{\delta}_{n}(X, X_i, Y_i), & {\rm otherwise}
\end{array}%
\right..
\end{eqnarray}
It is straightforward to verify that (\ref{randomizedrule}) is a valid probability distribution for every $(X, X_i, Y_i)\in{\mathcal{X}}\times{\mathcal{X}}\times{\mathcal{Y}}$. Therefore, together with (A), $\delta_{ni}$ is clearly specified by ${\bar{\delta}}_{ni}(x)$ and (\ref{randomizedrule}).  

Note, this formalism serves merely as a technical convenience and should not mask the simplicity of the agent decision rules.  In words, an agent will abstain from voting if $\bar{\delta}_{n}(X, X_i, Y_i)={\rm abstain}$; else, the agent flips a biased coin to send $1$ or $0$, with the bias determined by $\bar{\delta}_{n}(X, X_i, Y_i)$.  Though this formalism may appear restrictive since rules of this form do not allow randomized decisions to abstain, the results in this paper do not rely on this flexibility.

To emphasize, note that communication is constrained between the agents and the fusion center via the limited decision space ${\mathcal{S}}$ and as above, communication between agents is not allowed (the latter is a necessary precondition for observation (A)).  Consistent with the notation, we assume that the agents have knowledge of $n$, the number of agents in the ensemble.  Moreover, we assume that for each $n$, every agent has the same local decision rule;  i.e., the ensemble is homogenous in this sense.   An underlying assumption is that each agent is able to generate random numbers, independent of the rest of the network. 

Consistent with convention, we use $g_n(x)=g_n(x, \{\delta{_{ni}}\}_{i=1}^n):{\mathcal{X}}\times{\mathcal{S}}^n\rightarrow\{0,1\}$ to denote the central authority's fusion rule in the binary classification frameworks and similarly, we use $\hat{\eta}_n(x)=\hat{\eta}_n(x, \{\delta{_{ni}}\}_{i=1}^n\}):{\mathcal{X}}\times{\mathcal{S}}^n\rightarrow\IR$ to denote its fusion rule in the regression frameworks.   In defining fusion rules throughout the remainder of the paper, it will be convenient to denote the random set $I_V = I_V(X, D_n) \triangleq \{ i\in\{1,...,n\}:\, \delta_{ni} \neq {\rm abstain}\}$ as the set of agents that vote and hence, do not abstain.   To emphasize the central authority's primary role of aggregating the response of the network, we shall henceforth refer to this agent as a \emph{fusion center}.

Defining a loss function $l:{\mathcal{Y}}\times{\mathcal{Y}}\rightarrow\IR$, we seek ensembles that achieve the minimal expected loss.  In the binary classification setting, the criterion of interest is the probability of misclassification; we let $l(y,y^{\prime}) = 1_{\{y\neq y^{\prime}\}}$, the well-known zero-one loss.  The structure of the risk minimizing MAP decision rule is well-understood \cite{DevGyoLug96}; let $\delta_B:{\mathcal{X}}\rightarrow\{0,1\}$ denote this Bayes decision rule.  In regression settings, we consider the squared error criterion;  we let $l(y, y^{\prime}) = |y-y^{\prime}|^2$.    It is well known that the regression function
\begin{equation}\label{regressionfunction}
\eta(x) = {\mathbf{E}}\{Y\,| X=x\}
\end{equation}
 achieves the minimal expected loss in this case.  Throughout the remainder of the paper, we let $L^{\star}=\inf_{f}{\mathbf{E}}\{l(f(X), Y)\}$ denote the minimal expected loss.  Depending on whether we find ourselves in the binary classification or regression setting, it will be clear from the context whether $L^{\star}$ refers to the optimal (binary) Bayes risk or minimal mean squared error. 

In this work, we focus on the statistical property known as \emph{universal consistency} \cite{DevGyoLug96,GyoKohKrzWal02}, defined as follows.

\begin{defn}
Let $L_n={\mathbf{E}}\{l(f_n(X, D_n), Y)\,| D_n\}$. $\{f_n\}_{n=1}^{\infty}$ is said to be \emph{universally consistent} if $\mathbf{E}\{L_n\}\rightarrow L^{\star}$ for \emph{all} distributions ${\mathbf{P}}_{XY}$.
\end{defn}

This definition requires convergence in expectation and according to convention, defines \textit{weak} universal consistency. This notion is contrasted with \textit{strong} universal consistency where $L_n\rightarrow L^{\star}$ almost surely. Extending results of weak universal consistency to the strong sense has generally required the theory of large deviations, in particular McDiarmid's inequality \cite{DevGyoLug96}.   Though the focus in this paper is on the weaker sense, the results in this paper might be extended to strong universal consistency using similar techniques.  In particular, note that since consistency in distributed classification \textit{with abstention} can be reduced to Stone's Theorem, the extension to strong universal consistency follows immediately from standard results.  Further, the negative result  for distributed regression \textit{without abstention} automatically precludes consistency in the strong sense.  An extension for distributed classification without abstention and distributed regression with abstention may be possible under a refined analysis; the authors leave such analysis for future research.

\section{Distributed Classification with Abstention: Stone's Theorem}
In this section, we show that the universal consistency of distributed classification with abstention follows immediately from Stone's Theorem and the classical analysis of naive kernel classifiers.  To start, let us briefly recap the model.  Since we are in the classification framework, $\mathcal{Y}=\{0, 1\}$.  Suppose that for each $i\in\{1,...,n\}$, the training datum $(X_i, Y_i)\in D_n$ is received by a distinct member of a network of $n$ learning agents. When the fusion center observes a new observation $X\sim\mathbf{P}_X$,  it broadcasts the observation to the network in a request for information.  At this time, each of the learning agents can respond with at most one bit.  That is, each learning agent chooses whether or not to respond to the fusion center's request for information; and if an agent chooses to respond,  it sends either a $1$ or a $0$ based on a local decision algorithm.  Upon receiving the agents' responses, the fusion center combines the information to create an estimate of $Y$.

To answer the question of whether agent decision rules and fusion rules exist that result in a universally consistent ensemble, let us construct one natural choice.  With $B_{r_n}(x) = \{x^{\prime}\in\IR^d: \parallel x-x^{\prime}\parallel_2 \leq r_n\}$, let 
\begin{equation}\label{CWA-agent}
\bar{\delta}_{n}(x, X_i, Y_i)= \left\{%
\begin{array}{ll}
    Y_i, & {\rm if\,\,} X_i\in B_{r_n}(x)\\
     {\rm abstain}, & {\rm otherwise}
\end{array}%
\right.
\end{equation}
and
\begin{equation}
g_n(x)= \left\{%
\begin{array}{ll}
    1, & {\rm if\,\,}\sum_{i\in I_V}\delta_{ni} \geq \frac{1}{2}|I_V|\\
     0, & {\rm otherwise}
\end{array}%
\right.\,,
\end{equation}
so that $g_n(x)$ amounts to a majority vote fusion rule.  Recall from (\ref{randomizedrule}) that the agents' randomized responses are defined by $\bar{\delta}_n(\cdot)$.  In words, agents respond according to their training data label as long as the new observation $X$ is sufficiently close to their training observation $X_i$;  else, they abstain.  In this model with abstention, note that $\delta_{ni}$ is  $\{{\rm abstain}, 1,  0\}$-valued since $Y_i$ is binary valued and thus, the communications constraints are obeyed.

With this choice, it is straightforward to see that the net decision rule is equivalent to the plug-in kernel classifier rule with the naive kernel.  Indeed, 
\begin{equation}
g_n(x)= \left\{%
\begin{array}{ll}
    1, & {\rm if\,\,} \frac{\sum_{i=1}^{n}Y_i 1_{B_{r_n}(x)}(X_i)}{\sum_{i=1}^{n} 1_{B_{r_n}(x)}(X_i)}\geq \frac{1}{2}\\
     0, & {\rm otherwise}
\end{array}%
\right. .
\end{equation}
With this equivalence\footnote{Strictly speaking, this equality holds almost surely (a.s.), since the agents' responses are random variables.}, the universal consistency of the ensemble follows from Stone's Theorem applied to naive kernel classifiers.  With $L_n = \mathbf{P}\{g_n(X)\neq Y\,| D_n\}$, the probability of error of the ensemble conditioned on the random training data, we state this known result without proof as Theorem 1.

\begin{thm}{(\cite{DevGyoLug96})}
If $r_n\rightarrow 0$ and $(r_n)^d n\rightarrow\infty$ as $n\rightarrow\infty$, then $\mathbf{E}\{L_n\}\rightarrow L^{*}$ for all distributions $\mathbf{P}_{XY}$. \end{thm}

The kernel classifier with the naive kernel is somewhat unique amongst other frequently analyzed universally consistent classifiers in its relevance to the current model.  More general kernels (for instance, a Gaussian kernel) are not easily applicable as the real-valued weights do not naturally form a randomized decision rule.  Furthermore, nearest neighbor rules do not apply as a given agent's decision rule would then need to depend on the data observed by the other agents; such inter-agent communication is not allowed in the current model.

\section{Distributed Classification without Abstention}
As noted in the introduction, given the result of the previous section, it is natural to ask whether the communication constraints can be tightened.  Let us consider the second model in which the agents cannot choose to abstain.  In effect, each agent communicates one bit per decision.  Again, we consider the binary classification framework but as a technical convenience, adjust our notation so that  $\mathcal{Y}=\{+1, -1\}$ instead of the usual $\{0, 1\}$; also, agents now decide between sending $\pm 1$. The formalism introduced in Section II can be extended naturally to allow this slight modification; we allow $\delta_{ni}$ to be specified so that ${\mathbf{P}}\{\delta_{ni} = +1\,|X, X_i, Y_i\} = \bar{\delta}_{ni}(x, X_i, Y_i)$.  We again consider whether universally Bayes-risk consistent schemes exist for the ensemble.

Consider the randomized agent decision rule specified as follows:
\begin{equation}
\bar{\delta}_{ni}(x, X_i, Y_i)= \left\{%
\begin{array}{ll}
    \frac{1}{2}Y_i + \frac{1}{2}, & {\rm if\,\,} X_i\in B_{r_n}(x)\\
     \frac{1}{2}, & {\rm otherwise}
\end{array}%
\right..
\end{equation}
Recall from (\ref{randomizedrule}) that the agents' randomized responses are defined by $\bar{\delta}_n(\cdot)$.  Note that ${\mathbf{P}}\{\delta_{ni} = Y_i\,| X_i\in B_{r_n}(x)\} = 1$,  and thus, the agents respond according to their training data label if $x$ is sufficiently close to $X_i$.  Else, they simply ``guess", flipping an unbiased coin.  In this model without abstention, it is readily verified that each agent transmits one bit per decision as $\delta_{ni}$ is $\{\pm 1\}$-valued since {${\mathbf{P}}\{\delta_{ni}={\rm abstain}\} =  0$; thus, the communication constraints are obeyed.

A natural fusion rule is the majority vote. That is, the fusion center decides according to
\begin{equation}
g_n(x)= \left\{%
\begin{array}{ll}
    1, & {\rm if\,\,} \sum_{i=1}^{n}\delta_{ni} > 0\\
     -1, & {\rm otherwise}
\end{array}%
\right..
\end{equation}
As before, the natural performance metric for the ensemble is the probability of misclassification. Modifying our convention slightly, let $D_n=\{(X_i, Y_i, \delta_{ni})\}_{i=1}^{n}$ and define 
\begin{equation}\label{riskofrandomrule}
L_n = \mathbf{P}\{g_n(X)\neq Y\, | D_n\}.
\end{equation}
That is, $L_n$ is the conditional probability of error of the majority vote fusion rule conditioned on the randomness in agent training and agent decision rules.

\subsection{Main Result and Comments}
Theorem 2 specifies sufficient conditions for consistency for an ensemble using the described decision rules.

\begin{thm}
If $r_n\rightarrow 0$ and $(r_n)^d\sqrt{n}\rightarrow\infty$ as $n\rightarrow\infty$, then $\mathbf{E}\{L_n\}\rightarrow L^{*}$.
\end{thm}

Yet again, the conditions of the theorem strike a similarity with consistency results for kernel classifiers using the naive kernel.  Indeed, $r_n\rightarrow 0$ ensures that the bias of the classifier decays to zero.  However, $\{r_n\}_{n=1}^{\infty}$ must not decay too rapidly.  As the number of agents in the ensemble grows large, many, indeed most, of the agents will be ``guessing" for any given classification; in general, only a decaying fraction of the agents will respond with useful information.  In order to ensure that these informative bits can be heard through the noise introduced by the guessing agents, $(r_n)^d \sqrt{n}\rightarrow\infty$.  Note the difference between this result and that for naive kernel classifiers where $(r_n)^d n\rightarrow\infty$ assures a sufficient rate of convergence for $\{r_n\}_{n=1}^{\infty}$.

Notably, to prove this result, we show directly that the expected probability of misclassification converges to the Bayes rate.  This is unlike techniques commonly used to demonstrate the consistency of kernel classifiers, etc., which are so-called ``plug-in" classification rules.  These rules estimate the \textit{a posteriori} probabilities $\mathbf{P}\{Y=i\,|X\}$, $i=\pm 1$ and construct classifiers based on thresholding the estimate.  In this setting, it suffices to show that these estimates converge to the true probabilities in $L^p(\mathbf{P}_X)$.  However, for this model, we cannot estimate the \textit{a posteriori} probabilities and must resort to another proof technique; this foreshadows the negative result of Section VI.

With our choice of ``coin flipping" agent decision rules, one may be tempted to model the observations made by the fusion center as noise-corrupted labels from the training set and to thereby recover Theorem 2 from the literature on learning with noisy data.  However, note that since the fusion center does not have access to the agents' feature observations (i.e., $\{X_i\}_{i=1}^n$), the fusion rule cannot in general be modeled as a ``plug-in" classication rule as analyzed, for instance, in \cite{Lug92}.   Moreover, in contrast to the noise models considered in \cite{Lug92}, the agent decision rules here are statistically dependent on $X$ and are also dependent on $X_i$ in an atypical way: the noise statistics depend on $n$ and for particular $\mathbf{P}_{XY}$, one can show that as $n$ increases without bound, the probability that an agent guesses (a label is noisy) grows toward $1$.  These differences distinguish Theorem 2 from results in the literature on learning with noisy data.

\subsection{Proof of Theorem 2}
 \begin{proof}
Fix an arbitrary $\epsilon>0$. We will show that $\mathbf{E}\{L_n\} - L^{*}$ is less than $\epsilon$ for all  sufficiently large $n$.  Using the notation in (\ref{regressionfunction}), we write $\eta(x)=\mathbf{E}\{Y\, | X=x\}=\mathbf{P}\{Y=+1\,|X=x\} - \mathbf{P}\{Y=-1\,|X=x\}$ and define $A_{\epsilon}=\{x : |\eta(x)| > \frac{\epsilon}{2}\}$.  It follows that
 \begin{eqnarray}
 \nonumber \lefteqn{\mathbf{E}\{L_n\} - L^{*}}\\
 \nonumber & = & \mathbf{E}\Big{\{}\mathbf{P}\{g_n(X)\neq Y\,| D_n\}\Big{\}}  - \mathbf{P}\{\delta_B(X) \neq Y\}\\
 \nonumber& = & \mathbf{E}\Big{\{}\Big{(}\mathbf{P}\{g_n(X)\neq Y\,| D_n, X\} \\
 \label{yayaya} & &\,\,\, - \mathbf{P}\{\delta_B(X) \neq Y\, | X\}\Big{)}\cdot\Big{(}1_{A_{\epsilon}}(X) + 1_{\bar{A}_{\epsilon}}(X)\Big{)}\Big{\}},
 \end{eqnarray}
with the expectation in (\ref{yayaya}) being taken with respect to $X$ and $D_n$. Note that for all $x\in\bar{A}_{\epsilon}$, $\mathbf{P}\{\delta_B(X)\neq Y\,|X=x\} = \frac{1}{2}-\frac{|\eta(x)|}{2}\geq\frac{1}{2} - \frac{\epsilon}{4}$ and therefore, $\mathbf{P}\{g_n(X)\neq Y\,|D_n, X\}\leq 1 - \mathbf{P}\{\delta_B(X)\neq Y\,|X=x\} \leq \frac{1}{2} + \frac{\epsilon}{4}$.  Thus,
 \begin{eqnarray}
 \nonumber \lefteqn{\mathbf{E}\{L_n\} - L^{*}}\\
 \nonumber & \leq & \mathbf{E}\Big{\{}\Big{(}\mathbf{P}\{g_n(X)\neq Y\,| D_n, X\} - \\
 \nonumber & & \,\,\,\,\,\,\,\,\,\,\,\mathbf{P}\{\delta_B(X) \neq Y\, | X\}\Big{)}1_{A_{\epsilon}}(X) + \frac{\epsilon}{2} \Big{\}}\\
 \nonumber & \leq & \mathbf{P}\Big{\{}g_n(X)\neq \delta_B(X)\,\Big{|} X\in A_{\epsilon}\Big{\}}\mathbf{P}\Big{\{}A_{\epsilon}\Big{\}} + \frac{\epsilon}{2}.
\end{eqnarray}
Note that if $\mathbf{P}\{A_{\epsilon}\}=0$, then the proof is complete.  Let us proceed assuming $\mathbf{P}\{A_{\epsilon}\}>0$.  Clearly, it suffices to show that $\lim_{n\rightarrow\infty}\mathbf{P}\Big{\{}g_n(X)\neq \delta_B(X)\,\Big{|} X\in A_{\epsilon}\Big{\}}\leq\frac{\epsilon}{2}$.  Let us define the quantities
 \begin{equation}
 \nonumber m_n(x) = \mathbf{E}\{\eta(X)\delta_{ni}\,| X=x\}
 \end{equation}
 \begin{equation}
 \nonumber \sigma_n^2(x) = \mathbf{E}\{|\eta(X)\delta_{ni}-m_n(X)|^2\, | X=x\},
\end{equation}
with the expectation being taken over the random training data and the randomness introduced by the agent decision rules.  Respectively, $m_n(x)$ and $\sigma_n^2(x)$ can be interpreted as the mean and variance of the ``margin" of the agent response $\delta_{ni}$, conditioned on the observation $X$.  For large positive $m_n(x)$, the agents can be expected to respond ``confidently" (with large margin) according to the Bayes rule when asked to classify an object $x$.  For large $\sigma_n^2(x)$, the fusion center can expect to observe a large variance amongst the individual agent responses to $x$.  
 
 Fix any integer $k>0$.  Consider the sequence of sets indexed by $n$,
 \begin{equation}\nonumber
 B_{n,k} = \{x\in{\cal{X}} : m_n(x)n > k\sqrt{n}\sigma_n(x)\},
 \end{equation}
so that $x\in B_{n,k}$ if and only if $\frac{m_n(x)\sqrt{n}}{\sigma_n(x)}>k$.  We can interpret $B_{n,k}$ as the set of observations for which informed agents have a sufficiently strong signal compared with the noise of the guessing agents.  Then, 
\begin{eqnarray}
 \nonumber \lefteqn{\mathbf{P}\Big{\{}g_n(X)\neq \delta_B(X)\,\Big{|} X\in A_{\epsilon}\Big{\}}}\\% & = & 
 & = & \mathbf{P}\Big{\{} \eta(X)\sum_{i=1}^n \delta_{ni} < 0\,\Big{|} X\in A_{\epsilon}\Big{\}}\\
 \nonumber & = & \mathbf{P}\Big{\{} \eta(X)\sum_{i=1}^n \delta_{ni} < 0\,\Big{|} X\in A_{\epsilon}\cap B_{n,k}\Big{\}}\cdot\\
 \nonumber & & \,\,\,\mathbf{P}\{X\in B_{n,k}\,| X\in A_{\epsilon}\} \\
 \nonumber&  &  + \mathbf{P}\Big{\{} \eta(X)\sum_{i=1}^n \delta_{ni} < 0\,\Big{|} X\in A_{\epsilon}\cap \bar{B}_{n,k}\Big{\}}\cdot\\
\label{returntome}&& \mathbf{P}\{X\in\bar{B}_{n,k}\,| X\in A_{\epsilon}\}
 \end{eqnarray}
 Note that conditioned on $X$, $\eta(X)\sum_{i=1}^n \delta_{ni}$ is a sum of independent and identically distributed random variables with mean $m_n(X)$ and variance $\sigma_n^2(X)$.  Further, for $x\in B_{n,k}$, $\eta(x)\sum_{i=1}^n \delta_{ni} < 0$ implies $|\eta(x)\sum_{i=1}^n \delta_{ni} - m_n(x)n| > k\sqrt{n}\sigma_n^2(x)$.  Thus, it is straightforward to see that,
 \begin{eqnarray}
 \nonumber \lefteqn{\mathbf{P}\Big{\{} \eta(X)\sum_{i=1}^n \delta_{ni} < 0\,\Big{|} X\in A_{\epsilon}\cap B_{n,k}\Big{\}}}\\
 \nonumber & = & \mathbf{E}\Big{\{} \mathbf{P}\Big{\{} \eta(X)\sum_{i=1}^n \delta_{ni} < 0\,\Big{|} X \Big{\}}\, \Big{|} X\in A_{\epsilon}\cap B_{n,k} \Big{\}}\\
 \nonumber & \leq &  \mathbf{E}\Big{\{} \mathbf{P}\Big{\{} \Big{|}\eta(X)\sum_{i=1}^n \delta_{ni} - m_n(X)n\Big{|} > k\sqrt{n}\sigma_n(X)\,\Big{|} X \Big{\}}\\
 \nonumber & & \,\,\,\,\,\,\, \Big{|} X\in A_{\epsilon}\cap B_{n,k} \Big{\}}\\
\nonumber & \leq & \frac{1}{k^2}.
\end{eqnarray}
Here, the last statement follows from Markov's Inequality.  Choosing $k$ sufficiently large and returning to (\ref{returntome}), 
\begin{eqnarray}
 \nonumber \lefteqn{\mathbf{P}\Big{\{}g_n(X)\neq \delta_B(X)\,\Big{|} X\in A_{\epsilon}\Big{\}}} \\
 \nonumber& \leq & \frac{\epsilon}{2} + \mathbf{P}\{X\in\bar{B}_{n,k}\,| X\in A_{\epsilon}\}.
 \end{eqnarray}
 Now let us determine specific expressions for $m_n(x)$ and $\sigma_n^2(x)$, as dictated  by our choice of agent decision rules.  Clearly,
 \begin{eqnarray}
 \nonumber \lefteqn{m_n(x)} \\
 \nonumber & = & \eta(x)\mathbf{E}\{\delta_{ni}\, | X=x\}\\
 \nonumber & = & \eta(x){\mathbf{E}}\Big{\{}{\mathbf{E}}\{2\bar{\delta}_{ni}(X, X_i, Y_i) - 1\,| X, X_i, Y_i\}\,\Big{ |} X=x\Big{\}}\\
 \nonumber & = & \eta(x)\Big{(}0\cdot\mathbf{P}\{X_i\in \bar{B}_{r_n}(x)\} \\
 \nonumber & & \,\,\,\,\,\,\,\,\,\,\,\,\,\,\,\,\,\,+ \eta_n(x)\cdot\mathbf{P}\{X_i\in B_{r_n}(x)\} \Big{)}\\
 \nonumber & = & \eta(x)\eta_n(x)\int 1_{B_{r_n}(x)}(y)P_X(dy),
 \end{eqnarray}
 with $\eta_n(x) = \mathbf{E}\{\eta(X)\, | X\in B_{r_n}(x)\}$.
 Also,
 \begin{eqnarray}
 \nonumber \sigma_n^2(x) & = & \eta^2(x)\mathbf{E}\{|\delta_{ni}-\mathbf{E}\{\delta_{ni}\,|X=x\}|^2\, | X=x\}\\
 \nonumber & = & \eta^2(x)(1-\mathbf{E}\{\delta_{ni}\, | X=x\}^2).
 \end{eqnarray}
Thus,
\begin{eqnarray}
\nonumber\lefteqn{\mathbf{P}\{X\in\bar{B}_{n,k}\,\,| X\in A_{\epsilon}\}}\\
\nonumber & = & \mathbf{P}\{m_n(X)n < k\sqrt{n}\sigma_n(X)\,\,| X\in A_{\epsilon}\}\\
\nonumber & = & \mathbf{P}\Big{\{} \frac{\eta(X)\eta_n(X)\int 1_{B_{r_n}(X)}(y)P_X(dy)\sqrt{n}}{|\eta(X)|\sqrt{1-\mathbf{E}\{\delta_{ni}\, | X\}^2} }  < k \,\,\Big{|} X\in A_{\epsilon}\Big{\}}\\
\nonumber & = & \mathbf{P}\Big{\{}\Big{(} {\rm sgn}(\eta(X))\eta_n(X) \Big{)}\cdot\\
\nonumber & & \,\,\,\,\,\,\,\Big{(}\frac{\sqrt{n}\int 1_{B_{r_n}(X)}(y)P_X(dy) }{\sqrt{1-\mathbf{E}\{\delta_{ni}\, | X\}^2}} \Big{)} < k \Big{|} X\in A_{\epsilon} \Big{\}}.
 \end{eqnarray}
 For any $1\geq\gamma>0$, we have
 \begin{eqnarray}
\nonumber\lefteqn{\mathbf{P}\{X\in\bar{B}_{n,k}\,\,| X\in A_{\epsilon}\}}\\
\nonumber & \leq &  \mathbf{P}\Big{\{}\frac{\sqrt{n}}{\sqrt{1-\mathbf{E}\{\delta_{ni}\, | X\}^2}} \int 1_{B_{r_n}(X)}(y)P_X(dy)< k \Big{|}  \\
 \nonumber & & \hspace{.5in} X\in A_{\epsilon}, {\rm sgn}(\eta(X))\eta_n(X) > \gamma \Big{\}}\\
\label{twoparter} & & +\mathbf{P}\{{\rm sgn}(\eta(X))\eta_n(X) \leq\gamma\, | X\in A_{\epsilon}\}.
 \end{eqnarray}
 First, consider the second term.  With $\gamma=\frac{\epsilon}{4}$, it follows from our choice of $A_{\epsilon}$ that $\{{\rm sgn}(\eta(X))\eta_n(X) \leq\frac{\epsilon}{4}\}$ implies $\{|\eta(X)-\eta_n(X)|>\frac{\epsilon}{4}\}$.  Thus,
 \begin{eqnarray}
 \nonumber \lefteqn{\mathbf{P}\Big{\{}{\rm sgn}(\eta(X))\eta_n(X) \leq\frac{\epsilon}{4}\,\, \Big{|} X\in A_{\epsilon}\Big{\}}}\\
 \nonumber & \leq & \mathbf{P}\Big{\{}|\eta(X)-\eta_n(X)|>\frac{\epsilon}{4}\,\, \Big{|} X\in A_{\epsilon}\Big{\}}.
 \end{eqnarray}
 Since by technical Lemma 2 (see appendix), $\eta_n(X)\rightarrow\eta(X)$ in probability and by assumption $\mathbf{P}\{A_{\epsilon}\}>0$, it follows from technical Lemma 1 in the appendix that $\mathbf{P}\{{\rm sgn}(\eta(X))\eta_n(X) \leq\frac{\epsilon}{4}\, | X\in A_{\epsilon}\} \rightarrow 0$.
 
 Returning to (\ref{twoparter}) with $\gamma=\frac{\epsilon}{4}$, note that we have just demonstrated that\\  $\lim_{n\rightarrow\infty}\mathbf{P}\{{\rm sgn}(\eta(X))\eta_n(X) > \frac{\epsilon}{4}\}=1$.  Thus, to show that the first term converges to zero, by technical Lemma 1, it suffices to show that 
\begin{equation}
\frac{\sqrt{n}}{\sqrt{1-\mathbf{E}\{\delta_{ni}\, | X\}^2}} \int 1_{B_{r_n}(X)}(y)P_X(dy)\rightarrow\infty\,\,{\rm i.p.}
 \end{equation}
 Since $\frac{1}{\sqrt{1-\mathbf{E}\{\delta_{ni}\, | X\}^2}}\geq 1$, this follows from technical Lemma 3 in the appendix and the fact that $(r_n)^d\sqrt{n}\rightarrow\infty$.
 This completes the proof.
 \end{proof}

\section{Distributed Regression with Abstention}
We now turn our attention to distributed regression.  As in Section III, the model remains the same except that now $\mathcal{Y}=\IR$; that is, $Y$ is an $\IR$-valued random variable and likewise, agents receive real-valued training data labels, $Y_i$.  In this section, we consider communication with abstention.   With the aim of determining whether universally consistent ensembles can be constructed, let us devise candidate rules.  

For some as yet unspecified sequence of functions $T_n:\IR\rightarrow [0,1]$ and a sequence of real numbers $\{r_n\}_{n=1}^{\infty}$, consider the randomized agent decision rules specified as follows: 
\begin{equation}\label{agentdr}
\bar{\delta}_{ni}(x)= \left\{%
\begin{array}{ll}
    T_n(Y_i) & {\rm if\,\,} X_i\in B_{r_n}(x)\\
     {\rm abstain}, & {\rm otherwise}
\end{array}%
\right.,
\end{equation}
for $i=1,...,n$.    In words, the agents choose to vote only if $X_i$ is close enough to $X$; to vote, they flip a biased coin, with the bias determined by the size of the ensemble $n$ and  $Y_i$, via the function $T_n(\cdot)$.  In this model with abstention, note that $\delta_{ni}$ is  $\{{\rm abstain}, 1,  0\}$-valued and thus, the communication constraints are obeyed.

It is intuitively clear that $T_n(\cdot)$ should be designed so that the realization of random bit $\delta_{n,i}$ reveals information about the real-valued label $Y_i$ to the fusion center.  In particular, it is natural to ask whether any continuous bijective mapping $\IR$ to the interval $(0,1)$ would suffice in biasing the coin in a manner that is informative enough to provide universal consistency.  For example, one might chose $T_n(y) = T(y)=\frac{1}{1+e^{-y}}$ and consider agent decision rules of the form (\ref{agentdr}) in conjunction with a fusion rule like
\begin{eqnarray}\label{fusion}
\hat{\eta}_n(x) & = & T^{-1}\Big{(}\frac{\sum_{i\in I_V} \delta_{ni}}{|I_V|}\Big{)}.
\end{eqnarray}
Since agents have the flexibility to abstain, the fusion center can accurately estimate the average bias chosen by non-abstaining agents; the hope, then, is to determine the corresponding average label by inverting $T(\cdot)$. As observed in the proof, such a choice is not possible, in general,  since $T(\cdot)$ is nonlinear; such an approach introduces a systematic bias to the estimator and thereby prevents consistency.

If, however, $|Y|\leq B$ a.s. for some known $B>0$, it suffices to choose $T_n(\cdot)$ as the linear function mapping $[-B, B]$ to $[0,1]$.  Since in this case, $T_n^{-1}({\mathbf{E}}\{ \delta_{n,i}\,| X, X_i\}) = {\mathbf{E}}\{Y_i \,| X_i\}$, universal consistency then follows with trivial modifications to the proof of Stone's Theorem.  

This intuition leads us to a rule that captures consistency in the general case.  Though choices abound, we can choose $T_n$ to be piecewise linear.  In particular, let $\{c_n\}_{n=1}^{\infty}$ be an arbitrary sequence of real numbers such that $c_n\rightarrow\infty$  as $n\rightarrow\infty$ and choose,
\begin{equation}
T_n(Y_{i})= \left\{%
\begin{array}{ll}
    \frac{1}{2 c_n}Y_i + \frac{1}{2} &  |Y_i| \leq c_n\\
     \frac{1}{2}, & {\rm otherwise}
\end{array}%
\right.,
\end{equation}
and specify the fusion rule as
\begin{eqnarray}\label{fusion}
\hat{\eta}_n(x) & = & 2 c_n\Big{(}\frac{\sum_{i\in I_V} \delta_{ni}}{|I_V|} - \frac{1}{2}\Big{)}.
\end{eqnarray}
In words, the fusion center shifts and scales the average vote.  For appropriately chosen sequences $\{c_n\}_{n=1}^{\infty}$ and $\{r_n\}_{n=1}^{\infty}$, this ensemble is universally consistent, as proved by Theorem 3.

In particular, we will consider $L_n = \mathbf{E}\{|\hat{\eta}_n(X)-Y|^2\}$ with the expectation being taken over $X$, $D_n=\{(X_i, Y_i)\}_{i=1}^{n}$, and the randomness introduced in the agent decision rules.
\subsection{Main Result and Comments}
 Assuming an ensemble using the described decision rules, Theorem 3 specifies sufficient conditions for consistency.
 \begin{thm}
 Suppose $\mathbf{P}_{XY}$ is such that $\mathbf{P}_X$ is compactly supported and $\mathbf{E}\{Y^2\}<\infty$. If, as $n\rightarrow\infty$,
 \begin{enumerate}
\item $c_n\rightarrow\infty$,
\item $r_n\rightarrow 0$, and
\item $\frac{c_n^2}{n r_n^d}\rightarrow 0$, 
\end{enumerate}
then ${\mathbf{E}}\{L_n\}\rightarrow L^{\star}$. 
 \end{thm} 
 
More generally, the constraint regarding the compactness of $\mathbf{P}_{X}$ can be weakened.    As will be observed in the proof below, $\mathbf{P}_X$ must be such that when coupled with a bounded random variable $Y$, there is a known convergence rate of the variance term of the naive kernel classifier (under a standard i.i.d. sampling model).  $\{c_n\}_{n=1}^{\infty}$ should be chosen so that it grows at a rate slower than the rate at which the variance term decays.   Notably, to select $\{c_n\}_{n=1}^{\infty}$, one does not need to understand the convergence rate of the bias term, and this is why continuity conditions are not required;  the bias term will converge to zero universally as long as $c_n\rightarrow\infty$ and $r_n\rightarrow 0$ as $n\rightarrow\infty$.

In observing the response of the network, the fusion center sees $\delta_{ni}$ from those agents who have not abstained.  Since these random variables can be viewed as random quantizations or transformations of the labels in the training data, it is natural to ask whether the consistency of these rules follows as a special case of models for learning with noisy data. In this case, the underlying noise model would transform the label $Y_i$ to the set $\{0,1\}$ in a manner that would be statistically dependent on $X$, $X_i$, $Y_i$ itself and $n$.  Though it is possible to view the current question in this framework, to our knowledge such a highly structured noise model has not been considered in the literature.

Finally, those familiar with the classical statistical pattern recognition literature will find the style of proof very familiar; special care must be taken to demonstrate that the variance of the estimate does not decrease too slowly compared to  $\{c_n\}_{n=1}^{\infty}$ and to show that the bias introduced by the ``clipped" agent decision rules converges to zero.

\subsection{Proof of Theorem 3}
\begin{proof}
By standard orthogonality arguments \cite{GyoKohKrzWal02}, it suffices to show that ${\mathbf{E}}\{|\hat{\eta}_n(X) - \eta(X)|^2\}\rightarrow 0$ as $n\rightarrow 0$. 

Define $\bar{\eta}_n(x)\triangleq\mathbf{E}\{\delta_{ni}\,|X_i=x, \parallel X - X_i \parallel \leq r_n\}$. Proceeding in the traditional manner, note that by the standard inequality
\begin{equation}\label{sumofsquares}
 (a_1+ \cdots +a_k)^2\leq k (a_1^2 + \cdots + a_k^2),
\end{equation}
it follows that
\begin{eqnarray}
\nonumber \lefteqn{\mathbf{E}\{|\hat{\eta}_n(X)-\eta(X)|^2\}}\\
\nonumber & \leq &  2\mathbf{E}\Big{\{}\Big{|}2 c_n\Big{(}\frac{\sum_{i\in I_V} \delta_{ni}}{|I_V|} - \frac{1}{2}\Big{)} -  2 c_n\Big{(}\frac{\sum_{i\in I_V} \bar{\eta}_{n}(X_i)}{|I_V|} - \frac{1}{2}\Big{)}\Big{|}^2\Big{\}}\\
\nonumber & & + \,\, 2\mathbf{E}\Big{\{}\Big{|}2 c_n\Big{(}\frac{\sum_{i\in I_V} \bar{\eta}_n(X_i)}{|I_V|} - \frac{1}{2}\Big{)} - \eta(X)\Big{|}^2\Big{\}}\\
\nonumber & \triangleq & J_n + K_n.
\end{eqnarray}
Starting with the first term,
\begin{eqnarray}
\nonumber \lefteqn{J_n} \\
\nonumber & = & 8 c_n^2 \mathbf{E}\Big{\{}\Big{|}\frac{\sum_{i\in I_V}(\delta_{ni}-\bar{\eta}_n(X_i)) }{|I_V|}\Big{|}^2\Big{\}}\\
\nonumber & = &8 c_n^2 \mathbf{E}\Big{\{} \mathbf{E}\Big{\{}\frac{\sum_{i\in I_V}(\delta_{ni}-\bar{\eta}_n(X_i))^2}{|I_V|^2}\,\Big{|}X,X_1,...,X_n\Big{\}}\Big{\}}.
\end{eqnarray}
Here, the first equality follows from algebra;  the second follows after noting that for all $i\in I_V$, \newline${\mathbf{E}}\{\delta_{ni}\,|X, X_1,...,X_n|\} = \hat{\eta}_n(X_i)$ and canceling out cross-terms in the expansion of the squared sum in the numerator.   Note that conditioned on $X$ and $X_i$, $\delta_{ni}$ is Bernoulli with parameter $\bar{\eta}_n(X_i)$ for all $i\in I_V$. Thus, bounding the variance of a Bernoulli random variable, we continue above,   
\begin{eqnarray}
\nonumber & \leq & 2 c_n^2 \mathbf{E}\Big{\{}\frac{1}{|I_V|}1_{\{|I_V|>0\}}\Big{\}}.
\end{eqnarray}
Here we have applied the convention $\frac{0}{0}=0$. Conditioning on $X$ and applying technical Lemma 4 (see the appendix) to the binomial random variable $|I_V| = \sum_{i=1}^n 1_{\{X_i\in B_{r_n}(X)\}}$, it follows that,
\begin{eqnarray}
\label{ok2} J_n& \leq &2 c_n^2 \mathbf{E}\Big{\{} \frac{2}{n\mathbf{P}_{X_1}\{X_1\in B_{r_n}(X)\}} \Big{\}}.
\end{eqnarray} 
Here, for convenience, we have exploited the fact that $D_n$ is i.i.d. and reused the variable $X_1$.   Since $\mathbf{P}_X$ is compactly supported, the expectation in (\ref{ok2}) can be bounded by a term $O(\frac{1}{n r_n^d})$ using an argument typically used to demonstrate the consistency of kernel estimators \cite{GyoKohKrzWal02}.  For completeness, we include it here.

Since $S$, the support of $\mathbf{P}_X$, is compact, we can find $z_{1}$,...,$z_{M_n}$$\in\IR^d$ such that $S\subseteq \cup_{i}^{M_n} B_{r_n/2}(z_i)$ and $M_n\leq\frac{c_1}{r_n^d}$ for some constant $c_1$.  Thus,

\begin{eqnarray}
\nonumber \lefteqn{2 c_n^2 \mathbf{E}\Big{\{} \frac{2}{n\mathbf{P}_{X_1}\{X_1\in B_{r_n}(X)\}} \Big{\}}}\\
\nonumber & \leq & 4 c_n^2 \sum_{i=1}^{M_n}\mathbf{E}\Big{\{} \frac{1_{\{B_{r_n/2}(z_i)\}}(X)}{n\mathbf{P}_{X_1}\{X_1\in B_{r_n}(X)\}} \Big{\}} \\
\nonumber & \leq &  4 c_n^2 \sum_{i=1}^{M_n}\mathbf{E}\Big{\{} \frac{1_{\{B_{r_n/2}(z_i)\}}(X)}{n\mathbf{P}_{X_1}\{X_1\in B_{r_n/2}(z_i)\}} \Big{\}}\\
\nonumber & = & \frac{4 c_n^2 M_n}{n}\\
\nonumber & \leq & \frac{4 c_1 c_n^2}{n r_n^d}. 
\end{eqnarray} 
Finally, by condition (3) of Theorem 3, it follows that $J_n\rightarrow 0$.  Note that $J_n$ is essentially the variance of the estimator. Much of the work thus far has been the same as showing that in traditional i.i.d. sampling process settings, the variance of the naive kernel is universally bounded by a term $O(\frac{1}{n r_n^d})$ when $\mathbf{P}_X$ is compactly supported and $Y$ is bounded \cite{GyoKohKrzWal02}.  This observation is consistent with the comments above.

Now, let us consider $K_n$.  Fix $\epsilon>0$.  We will show that for all sufficiently large $n$, $K_n<\epsilon$. Let $\eta_{\epsilon}(x)$ be a bounded continuous function with bounded support such that $\mathbf{E}\{|\eta_{\epsilon}(X)-\eta(X)|^2\}\leq\frac{\epsilon}{12}$.  Since $\mathbf{E}\{Y^2\}<\infty$ implies that $\eta(x)\in L^2(\mathbf{P}_X)$, such a function is assured to exist; the set of bounded continuous functions with bounded support is dense in $L^2(\mu)$ for all probability measures $\mu$.  By (\ref{sumofsquares}),
\begin{eqnarray}
\nonumber K_n & \leq & 4\mathbf{E}\Big{\{}\Big{|}2 c_n\Big{(}\frac{\sum_{i\in I_V} \bar{\eta}_n(X_i)}{|I_V|} - \frac{1}{2}\Big{)} - \frac{\sum_{i\in I_V} \eta_{\epsilon}(X_i)}{|I_V|}\Big{|}^2\Big{\}}\\
\nonumber & & +\,\, 4\mathbf{E}\Big{\{}\Big{|}\frac{\sum_{i\in I_V} \eta_{\epsilon}(X_i)}{|I_V|} - \frac{\sum_{i\in I_V} \eta_{\epsilon}(X)}{|I_V|}\Big{|}^2\Big{\}}\\
\nonumber & &+\,\, 4\mathbf{E}\Big{\{}\Big{|}\frac{\sum_{i\in I_V} \eta_{\epsilon}(X)}{|I_V|}-\eta_{\epsilon}(X)\Big{|}^2 \Big{\}}\\
\nonumber & & +\,\, 4\mathbf{E}\{|\eta_{\epsilon}(X)-\eta(X)|^2\}\\
\nonumber & \triangleq & 4 (K_{n1} + K_{n2} + K_{n3} + K_{n4}).
\end{eqnarray}
First, consider $K_{n1}$.
\begin{eqnarray}
\nonumber \lefteqn{K_{n1}}\\
\nonumber & = & \mathbf{E}\Big{\{}\Big{|}\frac{\sum_{i\in I_V} (2 c_n (\bar{\eta}_n(X_i)-\frac{1}{2}) - \eta_{\epsilon}(X_i))}{|I_V|}1_{\{|I_V|>0\}} \\
\nonumber & & \,\,\,\,\,\,\,\,\,\,\,- c_n 1_{\{|I_V|=0\}}\Big{|}^2\Big{\}}\\
\nonumber & \leq & 2\mathbf{E}\Big{\{}\Big{|}\frac{\sum_{i\in I_V} (2 c_n (\bar{\eta}_n(X_i)-\frac{1}{2}) - \eta_{\epsilon}(X_i))\}}{|I_V|}1_{\{|I_V|>0\}}\Big{|}^2\Big{\}} \\
\nonumber & & \,\,+ 2\mathbf{E}\{c_n^2 1_{\{|I_V|=0\}}\},
\end{eqnarray}
with the equality following from algebra and the inequality from (\ref{sumofsquares}).  Then, noting that $|I_V| = \sum_{i=1}^n 1_{\{X_i\in B_{r_n}(X)\}}$ is binomial with parameter $P_{X_1}\{X_1\in B_{r_n}(X)\}$ when conditioned on $X$, we continue,
\begin{eqnarray}
\nonumber K_{n1}& \leq & 2\mathbf{E}\Big{\{}\Big{|}\frac{\sum_{i\in I_V} (2 c_n (\bar{\eta}_n(X_i)-\frac{1}{2}) - \eta_{\epsilon}(X_i))}{|I_V|}\Big{|}^2\Big{\}} \\
\nonumber & &+ 2\mathbf{E}\Big{\{}c_n^2\Big{(}1-\mathbf{P}_{X_1}\{X_1\in B_{r_n}(X)\}\Big{)}^n\Big{\}}\\
\nonumber & \leq & 2c\mathbf{E}\Big{\{}\Big{|}2c_n(\bar{\eta}_n(X)-\frac{1}{2})-\eta_{\epsilon}(X)\Big{|}^2\Big{\}} \\
\nonumber & & + 2\mathbf{E}\Big{\{}\frac{2 c_n^2}{n\mathbf{P}_{X_1}\{X_1\in B_{r_n}(X)\}}\Big{\}}.
\end{eqnarray}
Here, the second inequality follows for some constant $c$, in part by applying technical Lemma 5 and in part by noting $(1-x)^n\leq\exp(-nx)\leq\frac{1}{nx}$ for $0\leq x\leq 1$ and $n=1,2, \cdots$.  Continuing by applying (\ref{sumofsquares}), we have
\begin{eqnarray}
\nonumber K_{n1} & \leq & 2c\mathbf{E}\Big{\{}\Big{|}2c_n(\bar{\eta}_n(X)-\frac{1}{2})-\eta(X)\Big{|}^2\Big{\}} \\
\nonumber & & \,\,\,\,+ \mathbf{E}\{|\eta_{\epsilon}(X)-\eta(X)|^2\} \\
\nonumber & & \,\,\,\,+ \mathbf{E}\Big{\{}\frac{4 c_n^2}{n\mathbf{P}_{X_1}\{X_1\in B_{r_n}(X)\}}\Big{\}}.
\end{eqnarray}
For our specific choice of agent decision rules, note that $\bar{\eta}_n(x)={\mathbf{E}}\{T_n(Y)\,|X=x\} = \mathbf{E}\Big{\{}(\frac{1}{2 c_n}Y + \frac{1}{2})1_{\{|Y|\leq c_n\}} + \frac{1}{2}1_{\{|Y|>c_n\}}\,\Big{|} X=x\Big{\}}$.  Substituting this above and applying Jensen's inequality, we have
\begin{eqnarray}
\nonumber K_{n1} & \leq & 2c\mathbf{E}\Big{\{}\Big{|}\mathbf{E}\{Y 1_{\{|Y|>c_n\}}\,| X\}\Big{|}^2\Big{\}} + \frac{\epsilon}{12} \\
\nonumber & & \,\,\,\,\,+ \mathbf{E}\Big{\{}\frac{4 c_n^2}{n\mathbf{P}_{X_1}\{X_1\in B_{r_n}(X)\}}\Big{\}}\\
\nonumber & \leq & 2c\mathbf{E}\Big{\{}\mathbf{E}\{Y^2 1_{\{|Y|>c_n\}}\,| X\}\Big{\}} + \frac{\epsilon}{12} \\
\nonumber & & \,\,\,\,\,+ \mathbf{E}\Big{\{}\frac{4 c_n^2}{n\mathbf{P}_{X_1}\{X_1\in B_{r_n}(X)\}}\Big{\}}\\
 \nonumber& = & 2c\mathbf{E}\{Y^2 1_{\{|Y|>c_n\}}\} + \frac{\epsilon}{12}\\
\label{ok4} & & \,\,\,\,\, + \mathbf{E}\Big{\{}\frac{4 c_n^2}{n\mathbf{P}_{X_1}\{X_1\in B_{r_n}(X)\}}\Big{\}}.
\end{eqnarray}
Since $f_n(y)=y^2 1_{\{|y|>c_n\}}$ is a monotonically decreasing sequence of functions and $f_n(y)\rightarrow 0$ everywhere, then by the Monotone Convergence Theorem, the first term in (\ref{ok4}) converges to zero.  The third term in (\ref{ok4}) converges to zero by the same argument that was applied for $J_n$.  Thus, $\limsup_{n\rightarrow\infty}K_{n_1}\leq \frac{\epsilon}{12}$.

Observe that $\eta_{\epsilon}$ is uniformly continuous, since by construction, it is a bounded continuous function with bounded support.  Let $\delta>0$ be such that if $\parallel x-x^{\prime}\parallel<\delta$, then $|\eta_{\epsilon}(x)-\eta_{\epsilon}(x^{\prime})|\leq\sqrt{\frac{\epsilon}{12}}$.  Since $r_n\rightarrow 0$, for all sufficiently large $n$, $r_n<\delta$.  Thus, for all sufficiently large $n$,
\begin{eqnarray}
\nonumber K_{n2} & = & \mathbf{E}\Big{\{}\Big{|}\frac{\sum_{i\in I_V} \Big{(}\eta_{\epsilon}(X_i)-\eta_{\epsilon}(X)\Big{)}}{|I_V|}\Big{|}^2\Big{\}}\\
\nonumber & \leq & \frac{\epsilon}{12},
\end{eqnarray} 
since for all $i\in I_V$, $\parallel X_i - X \parallel \leq r_n$.  Next, consider $K_{n3}$.  We have
\begin{eqnarray}
\nonumber K_{n3} & = & \mathbf{E}\{\eta_{\epsilon}(X)^2 1_{\{|I_V|=0\}}\}\\
\nonumber & \leq& \sup_x(\eta_{\epsilon}(x)^2)\mathbf{E}\{1_{\{|I_V|=0\}}\}\\
\nonumber & \leq & \sup_x(\eta_{\epsilon}(x)^2)\mathbf{E}\Big{\{}\frac{2 c_n^2}{n\mathbf{P}_{X_1}\{X_1\in B_{r_n}(X)\}}\Big{\}},
\end{eqnarray}
in the usual way, as we see that $K_{n3}\rightarrow 0$.  Finally, $K_{n4}\leq\frac{\epsilon}{12}$ by our choice of $\eta_{\epsilon}(x)$.  Thus, 
\begin{eqnarray}
\nonumber \limsup_{n\rightarrow\infty} K_n & \leq &  4\Big{(}\frac{\epsilon}{12} + \frac{\epsilon}{12} + 0 + \frac{\epsilon}{12}\Big{)}\\
\nonumber & = & \epsilon.
\end{eqnarray}
Since $\epsilon$ was arbitrary, it is clear that $K_n$ converges to zero.
This completes the proof. \end{proof}

\section{Distributed Regression without Abstention}
Finally, let us consider the model for distributed regression without abstention.  Now, $\mathcal{Y}=\IR$; agents will receive real-valued training data labels $Y_i$.  However, when asked to respond with information, they will reply with either $0$ or $1$, as abstention is not an option.   

In this section, we first establish natural regularity conditions for candidate fusion rules and specify a reasonable class of agent decision rules. As an important negative result, we then demonstrate that for any agent decision rule within this class, there does not exist a regular fusion rule that is $L_2$ consistent for every distribution ${\mathbf{P}}_{XY}$.  This result establishes the impossibility of universal consistency in this model for distributed regression without abstention for a restricted, but reasonable class of decision rules.

To begin, consider the set of agent decision rules specified according to (\ref{randomizedrule}) for some $\bar{\delta}_n(\cdot)$.  In this model without abstention, we require that the implicit responses satisfy ${\mathbf{P}}\{\delta_{ni} = {\rm abstain}\} = 0$, but we impose no additional constraints on the agent decision rules.   With the formalism introduced in Section II, this assumption is equivalent to assuming $\{\bar{\delta}_n(\cdot)\}_{n=1}^{\infty}\subset {\mathcal{A}} = \{ \delta:\mathcal{X}\times{\mathcal{X}}\times\mathcal{Y}\rightarrow[0,1]\}$.

A fusion rule consists of a sequence of functions $\{\hat{\eta}_n\}_{n=1}^{\infty}$ mapping ${\mathcal{X}}\times{\mathcal{S}}^n$ to $\mathcal{Y}=\IR$.  Recall from Section II, we can regard ${\mathcal{S}}=\{1,0\}$ in this model without abstention.  To proceed, we require some regularity on $\{\hat{\eta}_n(\cdot)\}_{n=1}^{\infty}$.  Namely, let us consider only fusion rules that satisfy the following assumptions:
\begin{description}
\item [(A1)] $\hat{\eta}_n(x, \cdot)$ is permutation invariant for all $x\in{\mathcal{X}}$.  That is, for all $x\in{\mathcal{X}}$, any $b\in\{0,1\}^n$, and any permutation of $b$, $b^{\prime}\in\{0,1\}^n$, $\hat{\eta}_n(x, b) = \hat{\eta}_n(x, b^{\prime})$.

\item [(A2)] For every $x\in{\mathcal{X}}$, $\hat{\eta}_n(x, \cdot)$ is Lipschitz  in the average Hamming distance.  That is, there exists a constant $C$ such that 
\begin{equation}\label{Lipschitz}
|\hat{\eta}_n(x, b_1) - \hat{\eta}_n(x, b_2)| \leq C \frac{1}{n}\sum_{i=1} ^n |b_{1i} - b_{2i}|
\end{equation}
for every $b_1, b_2\in\{0,1\}^n$.
\end{description}

Once again, we will consider $L_n = \mathbf{E}\{|\hat{\eta}_n(X)-Y|^2\}$ with the expectation being taken over $X$, $D_n=\{(X_i, Y_i)\}_{i=1}^{n}$, and the randomness introduced in the agent decision rules.

\subsection{Main Result and Comments}
The following provides a negative result.
\begin{thm}
For every sequence of agent decision rules specified according to (\ref{randomizedrule}) with a point-wise convergent sequence of functions $\{\bar{\delta}_n(\cdot)\}_{n=1}^{\infty}\subset {\mathcal{A}}$, there is no fusion rule $\{\hat{\eta}_n(\cdot)\}_{n=1}^{\infty}$ satisfying assumptions (A1) and (A2) such that
\begin{equation}
\lim_{n\rightarrow\infty}\mathbf{E}\{L_n\}=L^{\star}
\end{equation}
for every distribution $\mathbf{P}_{XY}$ satisfying ${\mathbf{E}}\{Y^2\}< \infty$.  
\end{thm}

Note that there is nothing particularly special about the one bit regime and regression.  In fact, under the conditions of the theorem, universal consistency cannot be achieved in a multi-class classification problem with even three possible labels.  However, we consider regression as it illustrates the point nicely.

The restriction to distributions satisfying ${\mathbf{E}}\{Y^2\}< \infty$ actually strengthens this negative result, for without such a condition, Theorem 4 is trivial.  In the proof, a counter-example is derived where $Y$ is binary-valued, a much stronger case that also satisfies this condition.  

Further, the requirement that $\{\bar{\delta}_n(\cdot)\}_{n=1}^{\infty}$ be pointwise convergent is mild and is only a technical point in the proof.  Indeed, the result can be trivially extended to allow for weaker notions of convergence.

\subsection{Proof of Theorem 4}
The proof will proceed by specifying two random variables $(X, Y)$ and $(X^{\prime}, Y^{\prime})$ with $\eta(x)=\mathbf{E}\{Y\,| X=x\}\neq\mathbf{E}\{Y^{\prime}\,| X^{\prime}=x\}=\eta^{\prime}(x)$.  Asymptotically, however, the fusion center's estimate will be indifferent to whether the agents are trained with random data distributed according to $\mathbf{P}_{XY}$ or $\mathbf{P}_{X^{\prime}Y^{\prime}}$.  This observation will contradict universal consistency and complete the proof.

\begin{proof}
To start, fix a pointwise convergent sequence of functions $\{\bar{\delta}_n(\cdot)\}_{n=1}^{\infty}\subseteq {\mathcal{A}}$, arbitrary $x_0, x_1\in{\mathcal{X}}$, and distinct $y_0,y_1\in\IR$.  Let us specify a distribution $\mathbf{P}_{XY}$.  Let $\mathbf{P}_X\{x_0\}=q$, $\mathbf{P}_X\{x_1\}=1-q$, and $\mathbf{P}_{Y|X}\{Y=y_i|X=x_i\}=1$ for $i=0,1$.  Clearly, for this distribution $\eta(x_i)=y_i$ for $i=0,1$.

Suppose that the ensemble is trained with random data distributed according to ${\mathbf{P}}_{XY}$ and that the fusion center wishes to classify $X=x_0$. According to the model, after broadcasting $X$ to the agents, the fusion center will observe a random sequence of $n$ bits $\{\delta_{ni}\}_{i=1}^n$.  For all $i\in\{1,...,n\}$ and all $n$,
\begin{eqnarray}
\lefteqn{\mathbf{P}\{\delta_{ni}=1\,|X=x_0\}} \\
\nonumber& = & \bar{\delta}_n(x_0, x_0, y_0)q + \bar{\delta}_n(x_0, x_1, y_1)(1-q).
\end{eqnarray}
Now, let us define a sequence of auxiliary random variables, $\{(X_n^{\prime}, Y^{\prime})\}_{n=1}^{\infty}$, with distributions satisfying
\begin{eqnarray}
\nonumber \lefteqn{\mathbf{P}_{X_n^{\prime}}\{x_1\}} \\
\nonumber &= &\frac{\bar{\delta}_n(x_0, x_0, y_0)q + \bar{\delta}_n(x_0, x_1, y_1)(1-q) - \bar{\delta}_n(x_0, x_1, y_1)}{\bar{\delta}_n(x_0, x_0, y_1) - \bar{\delta}_n(x_0, x_1, y_0)}\end{eqnarray}
\begin{eqnarray}
\nonumber \mathbf{P}_{X_n^{\prime}}\{x_0\} &=&1 - \mathbf{P}_{X_n^{\prime}}\{x_1\}\\
\mathbf{P}_{Y^{\prime}|X_n^{\prime}}\{Y^{\prime}=y_{1-i}\,|X_n^{\prime}=x_i\}&=&1,\,\,\,i=0,1.  
\end{eqnarray}
Here, $\eta^{\prime}(x_i)=\mathbf{E}\{Y^{\prime}\,|X_n^{\prime}=x_i\} = y_{1-i}$.  Suppose that the ensemble were trained with random data distributed according to ${\mathbf{P}}_{X_n^{\prime}Y^{\prime}}$ and let $\{\delta_{ni}^{(n)}\}_{i=1}^n$ denote the random response variables of the agents.  Then, we have
\begin{eqnarray}
\nonumber \lefteqn{\mathbf{P}\{\delta_{ni}^{(n)}=1\,|X_{n}^{\prime}=x_0\}}  \\
\nonumber  &  = &  \frac{\bar{\delta}_n(x_0, x_0, y_1)}{\bar{\delta}_n(x_0, x_0, y_1) - \bar{\delta}_n(x_0, x_1, y_0)}\Big{(}\bar{\delta}_n(x_0, x_0, y_0)q \\
\nonumber & & \,\,\,\,\,\,\,\,+ \bar{\delta}_n(x_0, x_1, y_1)(1-q) - \bar{\delta}_n(x_0, x_1, y_1)\Big{)} \\
 \nonumber & &\, + \frac{\bar{\delta}_n(x_0, x_1, y_0)}{\bar{\delta}_n(x_0, x_0, y_1) - \bar{\delta}_n(x_0, x_1, y_0)}\Big{(}1-\bar{\delta}_n(x_0, x_0, y_0)q\\
\nonumber & &  + \bar{\delta}_n(x_0, x_1, y_1)(1-q) - \bar{\delta}_n(x_0, x_1, y_1)\Big{)}\\
\label{one3} & = & \mathbf{P}\{\delta_{ni}=1\,|X=x_0\},
\end{eqnarray} 
for all $n$.  Thus,  conditioned on the observation to be labeled by the ensemble $X$ (or $X_n^{\prime}$), the fusion center will observe an identical stochastic process regardless of whether the ensemble was trained with data distributed according to $\mathbf{P}_{XY}$  or $\mathbf{P}_{X_n^{\prime}Y^{\prime}}$ for any fixed $n$. Note, this observation is true despite the fact that $\eta(x)\neq\eta^{\prime}(x)$.

Finally, let $(X^{\prime}, Y^{\prime})$ be such that
\begin{eqnarray}
\label{one5} \mathbf{P}_{X^{\prime}}\{x_1\} &= & \lim_{n\rightarrow\infty} \mathbf{P}_{X_n^{\prime}}\{x_1\} \\
\nonumber \mathbf{P}_{X^{\prime}}\{x_0\} &=&1 - \mathbf{P}_{X^{\prime}}\{x_1\}\\
\nonumber \mathbf{P}_{Y^{\prime}|X^{\prime}}\{Y^{\prime}=y_{1-i}\,|X^{\prime}=x_i\}&=&1,\,\,\,i=0,1.  
\end{eqnarray}
Again, $\eta^{\prime}(x_i)=\mathbf{E}\{Y^{\prime}\,|X^{\prime}=x_i\}=y_{1-i}$. These limits are assured to exist by the assumption that  $\{\bar{\delta}_n(\cdot)\}_{n=1}^{\infty}$ is a pointwise converging sequence of functions.  Finally, let $\{\delta_{ni}^\prime\}_{i=1}^n$ denote the random response random variables for the ensemble agents trained with data distributed according to $\mathbf{P}_{X^{\prime}Y^{\prime}}$.

By standard orthogonality arguments \cite{GyoKohKrzWal02}, for the ensemble to be universally consistent, we must have both
\begin{equation}\label{one1}
\mathbf{E}\{|\hat{\eta}_n(X, \{\delta_{ni}\}_{i=1}^n)-\eta(X)|^2\}\rightarrow 0
\end{equation}
and
\begin{equation}\label{two1}
\mathbf{E}\{|\hat{\eta}_n(X^{\prime},\{\delta_{ni}^{\prime}\}_{i=1}^n)-\eta^{\prime}(X^{\prime})|^2\}\rightarrow 0.
\end{equation}
Let us assume that (\ref{one1}) holds; we now demonstrate that necessarily,
\begin{equation}\label{three1}
\mathbf{E}\{|\hat{\eta}_n(X^{\prime},\{\delta_{ni}^{\prime}\}_{i=1}^n )-\eta(X^{\prime})|^2\}\rightarrow 0.
\end{equation}
Since $\eta(x)\neq\eta^{\prime}(x)$, (\ref{three1}) contradicts (\ref{two1}) and the proposition of universal consistency.  To show (\ref{three1}), it suffices to focus on the $L^2$ risk conditioned on $X^{\prime}$, due to the convenient point-mass structure of ${\mathbf{P}}_{X^{\prime}}$.  To proceed, note that by (\ref{sumofsquares}), for any $b\in\{0,1\}^n$,
\begin{eqnarray}
\nonumber \lefteqn{{\mathbf{E}\{|\hat{\eta}_n(X^{\prime},\{\delta_{ni}^{\prime}\}_{i=1}^n)-\eta(X^{\prime})|^2\,|X^{\prime}=x_0\}}}\\
\nonumber & \leq & 2\mathbf{E}\{|\hat{\eta}_n(X^{\prime}, b)-\eta(X^{\prime})|^2\,|X^{\prime}=x_0\} \\
\nonumber & & \, + 2\mathbf{E}\{|\hat{\eta}_n(X^{\prime}, \{\delta_{ni}^{\prime}\}_{i=1}^n) - \hat{\eta}_n(X^{\prime}, b)|^2\,|X^{\prime}=x_0\}\\
\nonumber & \triangleq & 2T_1(b) + 2T_2(b).
 \end{eqnarray}
In particular, let us select $b\in\{0,1\}^n$ randomly such that the components are i.i.d. with $b_i\sim {\mathbf{P}}\{\delta_{ni} \, | X=x_0\}$ for all $i=1,...,n$. Note that if we can show that ${\mathbf{E}}_b\{T_1(b) + T_2(b)\}\rightarrow 0$, then the result holds by the probabilistic method. First consider  $T_1(b)$.  Note that we have
 \begin{eqnarray}
\nonumber {\mathbf{E}}_b\{T_1(b)\} & = & \mathbf{E}\{|\hat{\eta}_n(X^{\prime}, b)-\eta(X^{\prime})|^2\,|X^{\prime}=x_0\} \\
\nonumber & = & \mathbf{E}\{|\hat{\eta}_n(X, \{\delta_{ni}\}_{i=1}^n)-\eta(X)|^2\,|X=x_0\} , 
\end{eqnarray}
by our selection of $b$. Thus,  ${\mathbf{E}}_{b}\{T_1(b)\}$ must converge to zero by the assumption that (\ref{one1}) holds true.  Considering $T_2(b)$, note that
\begin{eqnarray}
\nonumber \lefteqn{{\mathbf{E}}_b\{T_2(b)\}}\\
\nonumber & = & \mathbf{E}\{|\hat{\eta}_n(X^{\prime}, b) - \hat{\eta}_n(X^{\prime}, \{\delta_{ni}^{\prime}\}_{i=1}^n)|^2\,|X^{\prime}=x_0\} \\
\nonumber & \leq & C^2 {\mathbf{E}} \Big{\{}\Big{|} \frac{1}{n}\sum_{i=1}^n  b_{i} - \frac{1}{n}\sum_{i=1}^n \delta_{ni}^{\prime}\Big{|}^2\,\Big{|} X^{\prime}=x_0\Big{\}} \\
\label{helloworld1} & \leq & 3C^2 {\mathbf{E}} \Big{\{}\Big{|} \frac{1}{n}\sum_{i=1}^n  b_{i} - {\mathbf{P}}\{\delta_{ni}=1\,| X=x_0\} \Big{|}^2\Big{\}}\\
\nonumber & & + 3C^2 {\mathbf{E}} \Big{\{}\Big{|} \frac{1}{n}\sum_{i=1}^n \delta_{ni}^{\prime} - {\mathbf{P}}\{\delta_{ni}^{\prime}=1\,| X^{\prime}=x_0\}\Big{|}^2\,\Big{|} X^{\prime}=x_0\Big{\}} \\
\label{helloworld2} & & \\
\label{helloworld3} & & +  3C^2 |{\mathbf{P}}\{\delta_{ni}=1\,| X=x_0\} -  {\mathbf{P}}\{\delta_{ni}^{\prime}=1\,| X^{\prime}=x_0\}|^2.
\end{eqnarray}
Here, the first inequality follows from assumptions (A1) and (A2) and the second inequality follows by (\ref{sumofsquares}).  Note that since $\{b_i\}_{i=1}^n$ is i.i.d. with $b_i\sim{\mathbf{P}}\{\delta_{ni}=1\,| X=x_0\}$,
  \begin{eqnarray}
  \nonumber 3C^2 {\mathbf{E}} \Big{\{}\Big{|} \frac{1}{n}\sum_{i=1}^n  b_{i} - {\mathbf{P}}\{\delta_{ni}=1\,| X=x_0\} \Big{|}^2\Big{\}} & \leq & \frac{3C^2}{4n},
  \end{eqnarray}
after bounding the variance of a binomial random variable; therefore, (\ref{helloworld1}) must converge to zero.  A similar argument can be applied to (\ref{helloworld2}).  Next, from (\ref{one3}),
\begin{eqnarray}
\nonumber \lefteqn{ |{\mathbf{P}}\{\delta_{ni}=1\,| X=x_0\} -  {\mathbf{P}}\{\delta_{ni}^{\prime}=1\,| X^{\prime}=x_0\}|^2}\\
\nonumber & =  |{\mathbf{P}}\{\delta_{ni}^{(n)}=1\,| X_n^{\prime}=x_0\} -  {\mathbf{P}}\{\delta_{ni}^{\prime}=1\,| X^{\prime}=x_0\}|^2.
\end{eqnarray}
Thus, (\ref{helloworld3}) must converge to zero by our design of $(X^{\prime}, Y^{\prime})$ in (\ref{one5}).  Finally,  we have demonstrated that (\ref{three1}) holds true; by the discussion above, this completes the proof.
\end{proof}
\section{Conclusions and Future Work}
Motivated by sensor networks and other distributed settings, this paper has presented several models for distributed learning.  The models differ from classical works in statistical pattern recognition by allocating observations of an i.i.d. sampling process to individual learning agents.  By limiting the ability of the agents to communicate, we constrain the amount of information available to the ensemble and to the fusion center for use in classification or regression.   This setting models a distributed environment and presents new questions to consider with regard to universal consistency.

Insofar as these models present a useful picture of distributed scenarios, this paper has answered several questions about whether or not the guarantees provided by Stone's Theorem in centralized environments hold in distributed settings. The models have demonstrated that when agents are allowed to communicate $\log_2(3)$ bits per decision, the ensemble can achieve universal consistency in both binary classification and regression frameworks in the limit as the number of agents increases without bound.  In the binary classification case, we have demonstrated this property as a special case of naive kernel classifiers.  In the regression case, we have shown this to hold true with randomized agent decision rules.  When investigating the necessity of these $\log_2(3)$ bits, we have found that in the binary classification framework only one bit per agent per classification was necessary for universal consistency, and the analysis provided an interesting comparison for naive kernel methods in the traditional framework.  For regression, we have established the impossibility of universal consistency in the one bit regime for a natural, but restricted class of candidate rules.  

With regard to future research in distributed learning, there are numerous directions of interest. As these results are useful only if they accurately depict some aspect of distributed environments, other perhaps more reflective models are important to consider.  In particular, the current models assume that a reliable physical layer exists where bits transmitted from the agents are guaranteed to arrive unperturbed at the fusion center.  Future research may consider richer model for this communication, perhaps within an information-theoretic (i.e., Shannon-theoretic) formalism. Further, the current models consider simplified network models where the fusion center communicates with agents via a broadcast medium and each agent has a direct, albeit limited, channel to the fusion center.  Future research may focus on network models that allow for inter-agent communication. Consistent with the spirit of sensor networks, we might allow agents to communicate locally amongst themselves (or perhaps, hierarchically) before coordinating a response to the fusion center.  In general, models of this form would weaken (A) in the discussion in Section II by allowing for correlated agent responses.    A related assumption in this work is that the underlying data is i.i.d. Extending the results to other sampling process is important since in many distributed applications, the data observed by the agents may be correlated.  In this vein, connections to results in statistical pattern recognition results under non-i.i.d. sampling processes would be interesting and important to resolve.

Finally, from a learning perspective, the questions we have considered in this paper have been focused on the statistical issue of universal consistency.  Though such a consideration seems to be one natural first step, other comparisons between centralized and distributed learning are essential, perhaps with respect to convergence rate and the finite data reality that exists in any practical system.  Such questions open the door for agents to receive multiple training examples and may demand more complicated local decision algorithms; in particular, it may be interesting to study local regularization strategies for agents in an ensemble.  Future work may explore these and other questions frequently explored in traditional, centralized learning systems, with the hope of further understanding the nature of distributed learning under communication constraints.

\appendix
This appendix includes important facts that are commonly used in the study of nonparametric statistics and are similarly applied in the proofs above.   Lemma 1 is a basic result from probability theory and is included for clarity. Lemma 2 follows from Theorem 23.2 and Lemma 23.6 in \cite{GyoKohKrzWal02} applied to the naive kernel.  The proof of Theorem 6.2 in \cite{DevGyoLug96} contains the fundamental steps needed to prove Lemma 3.  Lemma 4 can be found as Lemma 4.1 in \cite{GyoKohKrzWal02}.  Lemma 5 follows from arguments used in proving Theorem 5.1 in \cite{GyoKohKrzWal02} applied to the naive kernel.

\begin{lem}
Suppose $\{X_n\}_{n=1}^{\infty}$ is a sequence of random variables such that $X_n\rightarrow X$ in probability.  Then, for any sequence of  events $\{A_n\}_{n=1}^{\infty}$ with $\liminf\mathbf{P}\{A_n\}>0$,
\begin{equation}
\nonumber\mathbf{P}\{|X_n-X| > \epsilon \,| A_n\}\rightarrow 0.
\end{equation}
for all $\epsilon>0$.
 \end{lem}
 \begin{proof}
After noting that, 
\begin{eqnarray}
\nonumber \lefteqn{\mathbf{P}\{|X_n - X| > \epsilon\}}\\
\nonumber & = & \mathbf{P}\{|X_n - X| > \epsilon\,| A_n\}\mathbf{P}\{A_n\} \\
\nonumber & & \,\,\,\,\,\,+   \mathbf{P}\{|X_n - X| > \epsilon\,|\bar{A}_n\}\mathbf{P}\{\bar{A}_n\}\\
\nonumber & \geq & \mathbf{P}\{|X_n - X| > \epsilon\,|A_n\}\mathbf{P}\{A_n\},
\end{eqnarray}
the Lemma follows trivially from the fact that $\liminf\mathbf{P}\{A_n\}>0$ and $X_n\rightarrow X$ in probability.  The proof follows similarly if $X_n\rightarrow\infty$ in probability.
 \end{proof}
 
 \begin{lem}
Let $X\sim{\mathbf{P}}_X$ be an $\IR^d$-valued random variable and fix any function $f\in L({\mathbf{P}}_X)$.  For an arbitrary sequence of real numbers $\{r_n\}_{n=1}^{\infty}$, define a sequence of functions $f_n(x)=\mathbf{E}\{f(X)\,|X\in B_{r_n}(x)\}$.    If $r_n\rightarrow 0$, then $f_n(X)\rightarrow f(X)$ in probability.
 \end{lem}
 
 \begin{lem}
Let $X\sim{\mathbf{P}}_X$ be an $\IR^d$-valued random variable and define $\{r_n\}_{n=1}^{\infty}$ and $\{a_n\}_{n=1}^{\infty}$ as arbitrary sequences of real numbers such that $r_n\rightarrow 0$ and $a_n\rightarrow\infty$.   If $(r_n)^d a_n\rightarrow\infty$, then
  \begin{equation}
  \nonumber
a_n \int 1_{B_{r_n}(X)}(y)P_X(dy)\rightarrow\infty\,\,{\rm i.p.}
 \end{equation}
 \end{lem}
 
\begin{lem}  Suppose $B(n,p)$ is a binomially distributed random variable with parameters $n$ and $p$. Then,
\begin{eqnarray}
\nonumber \mathbf{E}\Big{\{}\frac{1}{B(n,p)}1_{\{B(n,p)>0\}}\Big{\}} & \leq & \frac{2}{(n+1)p}.
\end{eqnarray}
\end{lem}

\begin{lem} There is a constant $c$ such that for any measurable function $f$, any $\IR^d$-valued random variable $X$, and any sequence  $\{r_n\}_{n=1}^{\infty}$,
\begin{eqnarray}
\nonumber \mathbf{E}\Big{\{}\frac{\sum_{i=1}^n 1_{\{X_i\in B_{r_n}(X)\}}f(X_i)}{\sum_{i=1}^n 1_{\{X_i\in B_{r_n}(X)\}}}\Big{\}} & \leq & c\mathbf{E}\{f(X)\}
\end{eqnarray}
for all $n$.
\end{lem}
 
 \nocite{Kol75}
%\bibliography{CL04-053}

\bibliographystyle{IEEEtranS}
 
%\begin{biography}[{\includegraphics[width=1in,height=1.25in,clip,keepaspectratio]{mshell}}]{Michael Shell}
% where an .eps filename suffix will be assumed under latex, and a .pdf suffix
% will be assumed for pdflatex; or if you just want to reserve a space for
% a photo:

\begin{biographynophoto}{Sanjeev R. Kulkarni}
(M'91, SM'96, F'04) received his Ph.D. from M.I.T. in 1991. From 1985 to 1991 he was a Member of the Technical Staff at M.I.T. Lincoln Laboratory.  Since 1991, he has been with Princeton University where he is currently Professor of Electrical Engineering. He spent January 1996 as a research fellow at the Australian National University, 1998 with Susquehanna International Group, and summer 2001 with Flarion Technologies.

Prof. Kulkarni received an ARO Young Investigator Award in 1992, an NSF Young Investigator Award in 1994, and several teaching awards at Princeton University.  He has served as an Associate Editor for the IEEE Transactions on Information Theory.  Prof. Kulkarni's research interests include statistical pattern recognition, nonparametric estimation,
learning and adaptive systems, information theory, wireless networks, and image/video processing.
\end{biographynophoto}

\begin{biographynophoto}{H. Vincent Poor}
(SÕ72, MÕ77, SMÕ82, FÕ77) received the Ph.D. degree in EECS from Princeton University in 1977.  From 1977 until 1990, he was on the faculty of the University of Illinois at Urbana-Champaign. Since 1990 he has been on the faculty at Princeton, where he is the George Van Ness Lothrop Professor in Engineering. Dr. PoorÕs research interests are in the areas of statistical signal processing and its applications in wireless networks and related fields. Among his publications in these areas is the recent book Wireless Networks: Multiuser Detection in Cross-Layer Design (Springer: New York, NY, 2005).

Dr. Poor is a member of the National Academy of Engineering and is a Fellow of the American Academy of Arts and Sciences. He is also a Fellow of the Institute of Mathematical Statistics, the Optical Society of America, and other organizations.  In 1990, he served as President of the IEEE Information Theory Society, and he is currently serving as the Editor-in-Chief of these Transactions. Recent recognition of his work includes the Joint Paper Award of the IEEE Communications and Information Theory Societies (2001), the NSF DirectorÕs Award for Distinguished Teaching Scholars (2002), a Guggenheim Fellowship (2002-03), and the IEEE Education Medal (2005).
\end{biographynophoto}

% insert where needed to balance the two columns on the last page
%\newpage

\begin{biographynophoto}{Joel B. Predd}
received a BS in electrical engineering from Purdue University in 2001, and an MA degree in Electrical Engineering from Princeton University in 2003.  Currently, he is a Ph.D. candidate in the Information Sciences and Systems group at Princeton.  He spent the summer of 2004 visiting the Statistical Machine Learning Group at National ICT Australia in Canberra; he was a Summer Associate at the RAND Corporation during summer 2005.  His research interests include nonparametric statistics, statistical machine learning, and the psychology of human decision making, with applications to distributed decision making and signal processing.  He is a student member of the IEEE.
\end{biographynophoto}


%\vfill


\end{document}